\documentstyle[11pt,newpasp,twoside,epsf]{article}
\def\edcomment#1{\iffalse\marginpar{\raggedright\sl#1\/}\else\relax\fi}
\reversemarginpar

\begin{document}
\title{Simulations of the Intergalactic Medium}
 \author{Romeel Dav\'e}
\affil{Astronomy Department, the University of Arizona, 933 North Cherry Avenue, Tucson, AZ, USA 85721}

\begin{abstract}
I present a review by epoch of baryons in the intergalactic medium (IGM),
from the first star until today.  Recent observations indicate a
protracted period of reionization, suggesting multiple populations of
reionizers; detection of these $z\ga 6$ sources is a key goal that is 
now coming within reach.  The optical Lyman alpha forest ($2\la z\la 4$)
is well-described by the Fluctuating Gunn-Peterson Approximation, but
puzzling observations of associated metal lines and nearby galaxies 
may provide insights into galactic feedback processes.
IGM studies at $z\la 1.5$ are
progressing rapidly thanks to ultraviolet absorption line studies of
both Ly$\alpha$ absorbers and the warm-hot intergalactic medium.
Cosmological hydrodynamic simulations have played an integral part
in these advances, and have helped to shape our understanding
of the IGM at each epoch.  Working jointly, observations and theory
continue to expand our knowledge of the IGM as the earliest stages of
galaxy formation and the dominant reservoir of baryons at all redshifts.
\end{abstract}

\section{Introduction}

All baryons in the Universe were originally in the intergalactic medium (IGM).
Today, the majority of baryons remain there.  Understanding the evolution
of the IGM is therefore a critical component for a complete theory of how 
baryons evolve in the Universe, from the smooth Dark Ages to present-day
galaxies and large-scale structure.  In this review I will highlight
some recent work towards understanding the evolution of the IGM's 
various components, their physical state and connection to galaxies,
with particular emphasis on the contributions and insights provided by 
numerical simulations.

This review is divided into four sections.  \S2 presents a simulation-based
overview of the various baryonic components in the Universe.  \S3 discusses
some of the latest theoretical and observational results probing the 
epoch of reionization.  \S4 presents a short history of optical Ly$\alpha$
forest studies, and expounds on two of its more puzzling aspects:
The presence of metals in the diffuse IGM, and the effects of galaxies
on their surrounding gas.  \S5 highlights recent developments in probing the
$z\la 1.5$ IGM, including prospects for detecting the so-called missing baryons.
\S6 provides a short summary.

\section{The History of Baryons in the Universe}

Baryons in the Universe can be divided into four phases, according to their
density and temperature (e.g. Dav\'e et al. 1999): (1)
``Condensed": Stars and cold gas in galaxies.  This is the most readily
detectable baryonic phase.
(2) ``Hot": Gas bound in clusters.  These baryons are seen via their 
X-ray emission, which generally requires that they have $T\ga 10^7$K.
(3) ``Diffuse": Low-density gas whose dynamics are governed by the
underlying dark matter distribution.  This phase gives rise to the
majority of Ly$\alpha$ absorbers seen in distant quasar spectra.
(4) ``Warm-Hot": Mildly shock-heated, uncollapsed gas with $10^5\la T\la 10^7$K.
This gas is difficult to detect in absorption due to its high ionization
state, and difficult to detect in emission due to its low density.
Hence these are the ``missing baryons" that have drawn considerable
attention, as we will discuss in \S5.

\begin{figure}
\plotfiddle{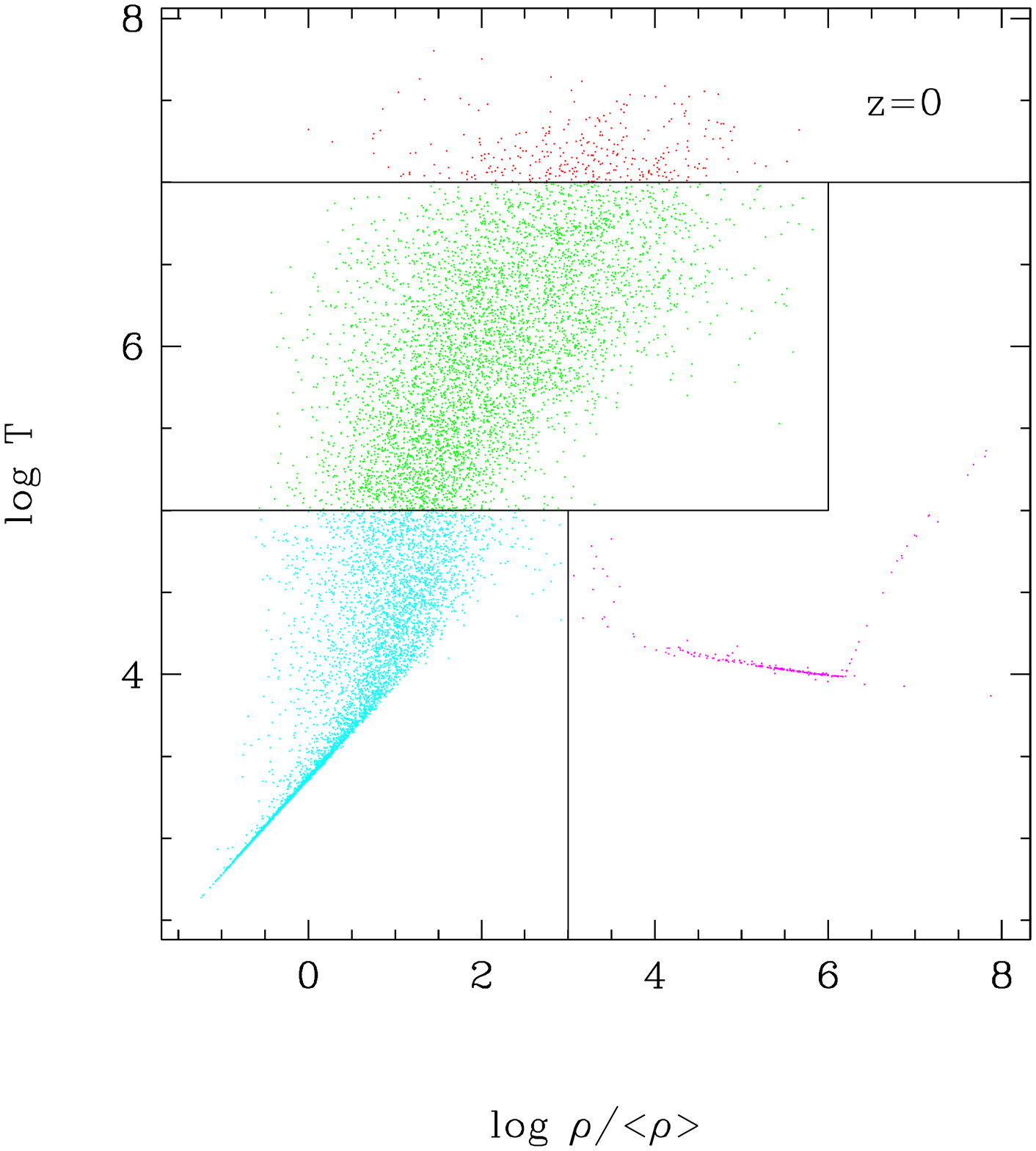}{1in}{0}{30}{30}{-190}{-135}
\plotfiddle{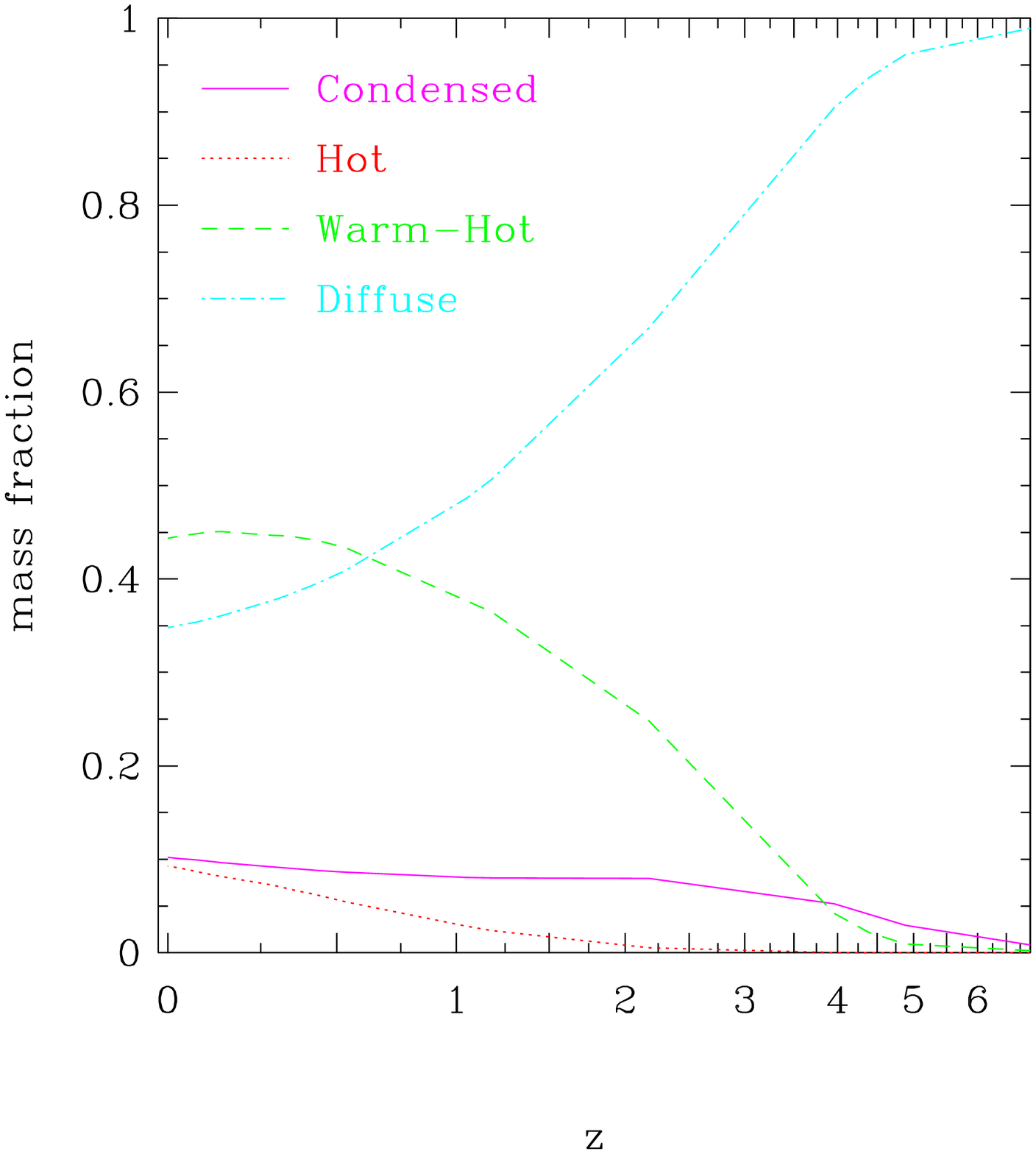}{1in}{0}{30}{30}{0}{-50}
\caption{{\it Left panel:} The cosmic phase diagram at $z=0$ from a 
$\Lambda$CDM simulation by Springel \& Hernquist (2003).
Phases described in text are demarcated.
{\it Right panel:} The evolution of the mass fraction of baryons in each phase from
$z=7\rightarrow 0$.}
\end{figure}

Figure~1 (left panel) shows the $z=0$ cosmological phase diagram, i.e. a
density-temperature plot of gas in a representative volume of the
Universe, drawn from a TreeSPH simulation by Springel \& Hernquist (2003,
their 100~Mpc/h G4 run).
They use kinetic feedback and a multi-phase ISM model to obtain a
converged star formation history that reproduces the observed 
stellar density at $z=0$.  Divisions between the four phases above
are indicated.  At the lowest densities, the diffuse phase
shows a tight density-temperature relation (discussed in \S4.2).
Moving to mildly overdense regions allows gas to compress and shock heat,
producing a scatter in temperature that grows with density through the
warm-hot phase.  As gas falls 
into the largest bound systems it virializes at the halo temperature and
enters the hot phase.  At larger densities, 
the cooling time becomes short and gas is able to cool into the
condensed phase and form stars.  The upturn at $\rho/\bar\rho>10^6$
arises from the details of feedback implementation in this simulation,
and the floor at $\sim 10^4$K in the condensed region arises because
no molecular hydrogen or metal line cooling is included in this run.

The right panel of Figure~1 shows the mass fraction in each phase
from $z\sim 7\rightarrow 0$.  At $z\ga 3$, virtually all baryons are in
the diffuse phase.  As structure forms, intergalactic gas in the filaments
surrounding galaxies is able to shock-heat, and the warm-hot phase 
begins to grow.  The condensed phase increases rapidly up to $z\sim 2$,
then the growth slows considerably, following the observed cosmic
star formation history.  Cluster formation governs the hot phase;
this becomes significant only at late times when large potential
wells are able to form.  By the present day, $\sim 45\%$ of the baryons
reside in the warm-hot phase.
This conclusion is broadly consistent with that from 
a number of other cosmological hydrodynamic simulations, using a 
variety of numerical techniques, resolutions, and input microphysics
(Dav\'e et al. 2001).  Thus these predictions for the evolution of
baryons are generic to CDM models, and are driven primarily by the
gravitational growth of structure (Valageas \& Silk 2001).

\section{The Reionization Epoch}

\subsection{The First Stars}

The IGM came into being, by definition, when the first galaxies formed.
At this time, the Universe was essentially metal- and dust-free,
with (presumably) negligible magnetic fields.  Thus the first star
formed in a relatively simple environment, from the gravitational
collapse of a primordial gas cloud (Tegmark et al. 1997).  Ultra-high resolution simulations
by Abel, Bryan \& Norman (2002) and Bromm, Coppi \& Larsen (2002)
show that a star produced in this environment is likely to be
a VMS (very massive star, $M\ga 100M_\odot$), since the lack of
H$_2$ and metal cooling agents results in a large Jeans mass.

The calculation of the second star is significantly more complex.  Due
to the exceptionally biased nature of the first collapsed object
(an $\sim 8\sigma$ perturbation), the second star would be expected
to form close to the first star.  However, the first star emits a
copious amount of ultraviolet radiation during its $\la 3$~million
year lifetime, which can sufficiently heat surrounding gas so as to inhibit
subsequent collapse.  If the collapse is inhibited beyond the lifetime
of the first star, then the resulting supernova may have a dramatic
impact on the second star's gas cloud.  VMS's were originally believed
to collapse directly to black holes with no ejectae; however, 
calculations by Heger \& Woosley (2002) show that in a specific stellar
mass range ($\approx 140-260M_\odot$), the resulting pair instability supernova
can expel a large fraction of its material into the surrounding IGM.
Not only would the mechanical wind energy evacuate surrounding gas,
but the accompanying metals would pollute the IGM.  If the pollution
reaches a modest value of $\sim 10^{-4}Z_\odot$, then subsequent
star formation from that gas will not be VMS's but will instead follow
a Population II IMF (Schneider et al. 2003).

From these considerations, it might appear that the first few massive stars
would serve to quickly pollute the IGM and transition to a lower-mass
IMF incapable of reionizing the Universe until late times.  This cannot
be the whole story, though, since
WMAP observations
of the temperature-polarization cross correlation
indicate a Thomson optical depth of $\tau_e=0.17\pm 0.04$, implying
a high reionization redshift of $\approx 11-30$ (95\% confidence; Kogut et al. 2003).
While the error bars are still large (and will improve with accumulating
WMAP data), if taken at face value they suggest a sizeable population
of efficiently-emitting VMS's in place at early times (Cen 2003b).

\subsection{The Minihalo Epoch}

The epoch of minihalos, i.e.
bound systems with virial temperatures below $10^4$K that
can only cool via molecular hydrogen line emission in a primordial gas, is
a crucial one for the reionization history of the Universe.  Minihalos
are so ubiquitous at $z\sim 10-15$ that they are expected to have a
covering fraction of unity around all halos having $T_{virial}>10^4$K
that are likely to be ionizing sources, thereby providing a large
sink for reionizing photons (Shapiro, Iliev \& Raga 2003).  However,
if minihalos are able to form stars themselves, they may provide a large source
of reionizing photons instead.  H$_2$ production in primordial gas
is insufficient to rapidly cool minihalos, but the presence of 
metals or X-rays that can
stimulate H$_2$ production could alter this scenario (Haiman, Abel \& Rees 
2000; though see Venkatesan, Giroux \& Shull 2001).  An early
population of black holes (Madau et al. 2003) may be a candidate 
for producing such hard X-rays.  However, Oh \& Haiman (2003) argue that
even if a minihalo forms an initial burst of stars, the resulting
increase in entropy in the relic \ion{H}{II} region would prevent
further gas collapse; thus they argue that minihalos cannot 
contribute significantly to $\tau_e$ measured
by WMAP.  Given the conflicting theoretical predictions, direct observations
of minihalos are highly desireable.  Prospects include seeing
them in 21cm emission with LoFAR or ALMA (Iliev et al. 2002, 2003),
or in 21cm absorption to background
sources (Furlanetto \& Loeb 2003) if such sources exist.

Since the Universe at $z\ga 10$ is sufficiently dense to recombine
after being initially ionized, and feedback from the first generation of 
stars may suppress subsequent star formation, it is possible that the Universe 
underwent two epochs of reionization (Cen 2003; Wyithe \& Loeb 2003a).  
The first would occur at $z\ga 15$ from VMS's (Oh et al. 2001; 
Wyithe \& Loeb 2003b), and the second around $z\sim 6$ from 
Population II stars in protogalaxies.  This would reconcile the WMAP
results with observations of a sharp rise in the IGM neutral fraction
seen in spectra of $z\ga 5.7$ Sloan Digital Sky Survey quasars (White
et al. 2003).  Sloan data also shows that the numbers of $z\sim 6$
quasars are too small to provide the necessary photons to perform this
final stage of reionization (Fan et al. 2003), thus star forming regions in
protogalaxies are the likely culprit.  An observational signature of
multiple reionizations may lie in 21cm emission from the IGM (Furlanetto,
Sokasian \& Hernquist 2003).

\subsection{Detecting the Reionizers at $z\ga 6$}

Observing the sources responsible for the final reionization of the Universe is
one science driver
for the James Webb Space Telescope (JWST), scheduled to lauch
in 2011.  Broad-band color selection can be used to identify candidates,
but detailed spectroscopic follow-up will be challenging from space,
and may require novel approaches in ground-based infrared spectroscopy.

\begin{figure}
\plottwo{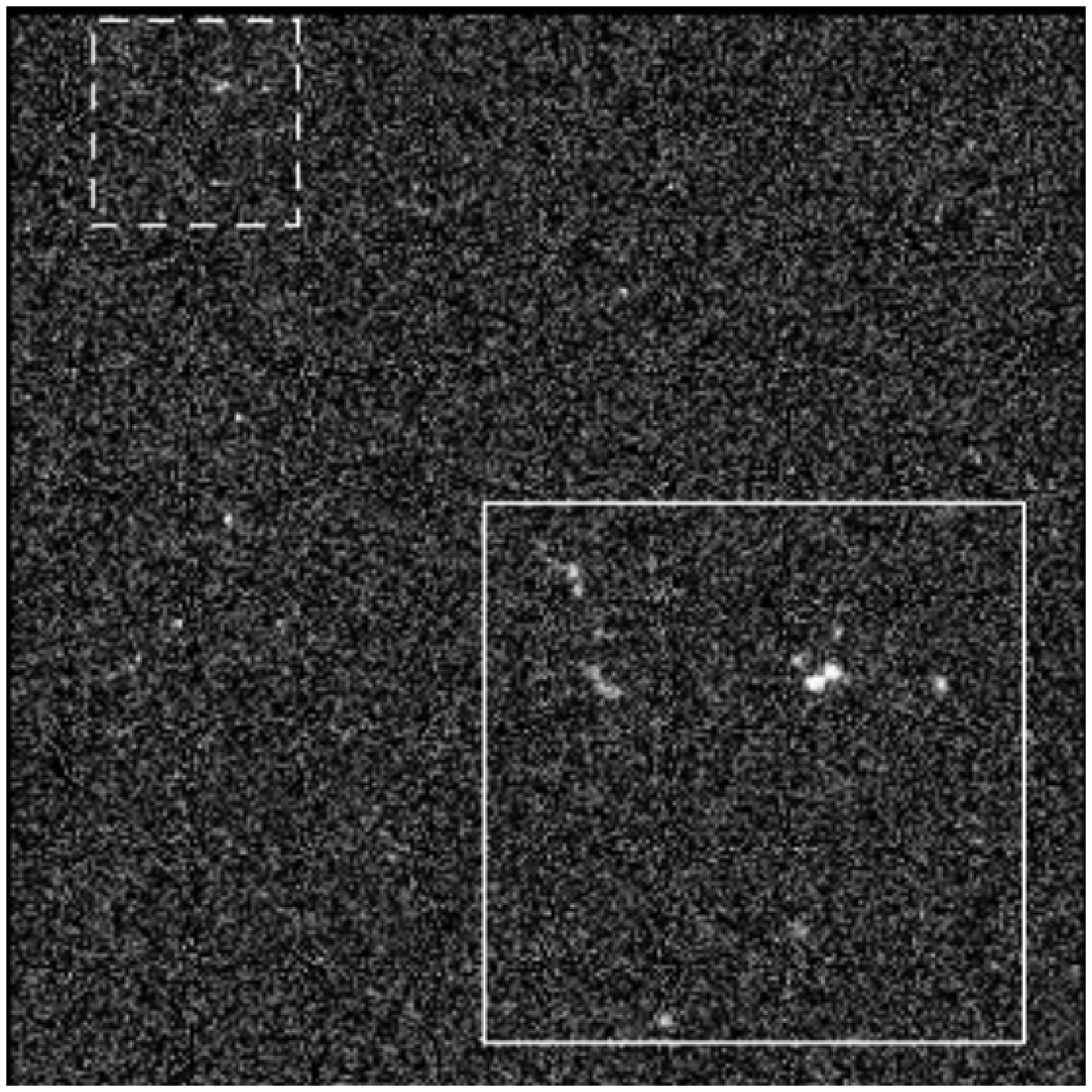}{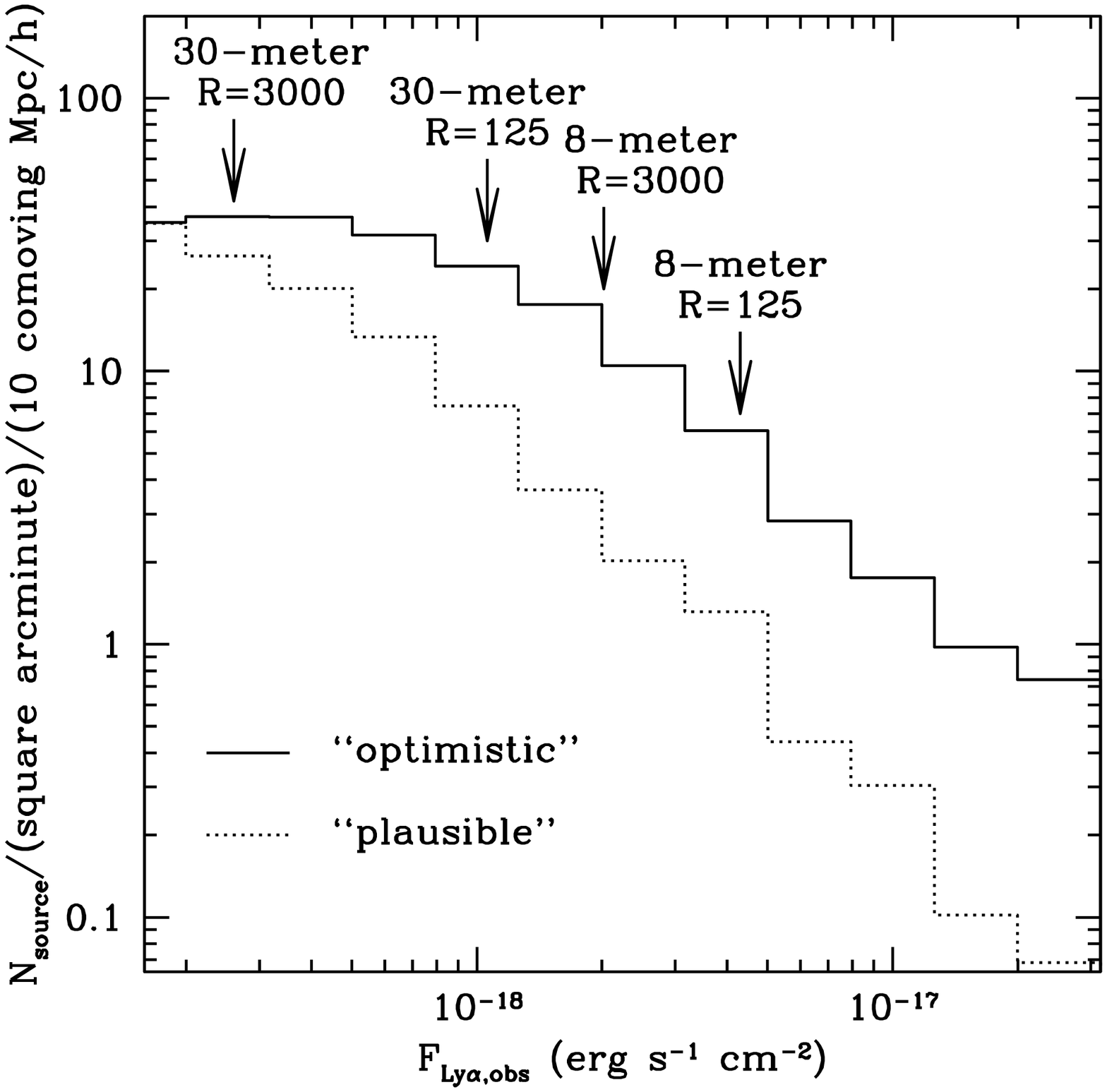}
\caption{{\it Left panel:} 32-hour ``observation" with 8m telescope 
of Ly$\alpha$ emission from $z=8.2$ galaxies in a $\Lambda$CDM simulation.  
Square inset shows a close-up of dashed box, the most heavily populated region.
{\it Right panel:} Ly$\alpha$ luminosity function for $z=8.2$ galaxies
in two IMF/$f_{\rm esc}$ scenarios , with detection limits for 8m and
30m telescopes with two narrow-band filters.  Figures from Barton 
et al. (2003).}
\end{figure}

Perhaps surprisingly, it may be more efficient to search for $z\ga 6$
star formation using ground-based narrow band searches for Ly$\alpha$
emission.  Cowie (priv. comm.) has confirmed several $z\approx 6.6$
sources this way, so Ly$\alpha$ emission is in fact detectable prior to
the end of reionization.  Barton et al. (2003) used a hydro simulation
to show that with optimistic (but not outrageous) assumptions
regarding the IMF and escape fractions, it may be possible to detect
pre-reionization objects with existing 8-10m telescopes, using an
$R\ga 100$ narrow-band filter in a night sky window corresponding to
Ly$\alpha$ emission at $z=8.2$.  Figure 2 (left panel) shows a simulated
32-hour observation with a VMS-based IMF and favorable escape parameters
from Santos (2003).  These include having sources blow a local ionization
bubble (Haiman 2003), having the emission be centered redwards of the
protogalaxy's redshift, and no dust attenuation.  Under these (admittedly
fortuitous) conditions, a fair number of sources would be detectable.

While detection with an 8m telescope will require some cooperation from
nature, detection with a 20m+ class telescope would be straightforward
even for fairly conventional IMFs and escape parameters.  The right panel
of Figure~2 shows the Ly$\alpha$ luminosity function predicted from the
Q5 simulation of Springel \& Hernquist (2003), having $2\times 324^3$
particles and a 10 Mpc/h (comoving) box size.  For a mildly top-heavy
IMF as might be expected for Population II stars, the Ly$\alpha$
luminosities would be $\sim 6\times$ lower than the ``plausible"
model.  This would preclude detection with an 8m class telescope,
at least without a large-area survey, but could be seen with a 
20-30m class telescope.  Such a telescope would also be required
for spectroscopic follow-up of JWST broad-band selected high redshift galaxies,
which is crucial for constraining their physical properties.

\section{The Optical Ly$\alpha$ Forest}

\subsection{History of Cloud Models}

Redshifted intervening weak neutral hydrogen absorption, i.e. the
``Ly$\alpha$ forest", was first detected in quasar spectra in the 1960's (e.g. Schmidt 1965;
Lynds \& Stockton 1966).  Their cosmological origin was clearly
demonstrated by Sargent et al. (1980), who also found a substantial 
number of associated \ion{C}{IV} absorbers.  Models for absorption
systems trace back to Gunn \& Peterson (1965) and Bahcall \& Salpeter (1965),
who forwarded the idea of a smooth, highly ionized IGM causing a
uniform ``Gunn-Peterson trough" (as it was later called), punctuated
by discrete clouds of neutral hydrogen.  This model persisted, in
various forms, into the 1990's, with individual cloud confinement
provided by ambient pressure from a hot IGM (Sargent et al. 1980;
Ikeuchi \& Ostriker 1986), gravity (Melott 1980; Black 1981), CDM
minihalos (Rees 1986; Ikeuchi 1986), and combinations thereof
(Petitjean et al. 1993).  With ever-improving observations, it became
clear that no model could fit Ly$\alpha$ data without 
a great deal of uncomfortable tuning.  The advent of Keck's HIRES
spectrograph, together with new insights from hydro simulations,
paved the way for the final demise of ``cloud" models.

\subsection{The Fluctuating Gunn-Peterson Model}

In the early 1990's, an alternative origin for Ly$\alpha$
absorbers was being formulated from rapidly advancing cosmological
hydrodynamic simulations.  It was noted that the Cosmic Web must
necessarily contain gas ionized by a metagalactic flux from quasars.
The resulting \ion{H}{I} absorption along any line of sight had
statistical properties remarkably similar to that of observed
Ly$\alpha$ systems (e.g. Cen et al. 1994; Zhang, Anninos \& Norman 1995).
In particular, the column density distribution (i.e. the number
of systems per unit column density per unit redshift) from such
models naturally reproduced the observed power law over 
$\sim 10$ orders of magnitude in $N_{\rm HI}$ (Hernquist et al. 1996;
Dav\'e et al. 1997).

From these simulations a new paradigm arose for the origin of the Ly$\alpha$
forest.  Instead of arising in individual clouds, Ly$\alpha$ absorption 
is analogous to a Gunn-Peterson trough that
fluctuates as it passes through over- and under-dense regions of the
Universe.  The absorbing gas is expected to have a tight density-temperature
relation set by the balance between adiabatic cooling due to Hubble
expansion and photoionization heating by the metagalactic UV background
(Hui \& Gnedin 1997).  
This ``equation of state" of the IGM can be
constrained observationally (Schaye et al. 2000; Choudhury, Padmanabhan, \& Srianand 2001).  
Thus a given density corresponds to a given
temperature, which corresponds to a given ionization fraction, which in turn
corresponds to a given \ion{H}{I} optical depth.  The simulation paradigm
can thus be encapsulated in the ``Fluctuating Gunn-Peterson Approximation"
(FGPA; Croft et al. 1998), in which $\tau_{\rm HI}\propto \rho^{1.6}$.
The density $\rho$ is that of either the dark matter or the gas, since
they trace each other quite well in these low-density, gravitationally-dominated
systems.  The constant of proportionality is inversely related to the \ion{H}{I}
photoionization rate, and also depends on the IGM temperature and
cosmological parameters (see Croft et al. for precise form).

The FGPA has proven to be a valuable tool for interpreting Ly$\alpha$
forest data.  For instance, according to the FGPA, the mean baryon 
density $\Omega_b$ can be simply related to the mean optical depth of
\ion{H}{I} absorption.  Rauch et al. (1997) measured $\bar\tau_{\rm HI}$,
and together with a minimal estimate of the \ion{H}{I} photoionization
rate from quasars, determined $\Omega_b\ga 0.017 h^{-2}$.  At the time,
this determination was considered quite high, though it agreed with
initial results from [D/H] studies (Tytler, Fan \& Burles 1996).  
WMAP results (Spergel et al. 2003) have now confirmed the Rauch et al. value.
Another application of the FGPA is to interpret metal lines observations
(discussed in \S4.3), using the conversion between $\tau_{\rm HI}$ and
$(\rho,T)$ to obtain accurate, pixel-by-pixel ionization corrections.

A more ambitious application of the FGPA, and the one for which it was
originally proposed, is the reconstruction of the matter power spectrum
(Croft et al. 1998).  Since the optical depth is a one-dimensional,
nonlinear map of the underlying dark matter distribution, it can be inverted
to yield the matter fluctuation spectrum.  The cosmological parameters
derived from the latest such analysis at $\bar z=2.7$ (Croft et al. 2002)
are consistent with the WMAP results at $z=1089$ (Spergel et al. 2003)
and large-scale structure studies at $z\la 0.5$, but uniquely probes
an intermediate redshift regime.  WMAP and the Croft et al. data together
show a mild preference for a running tilt of the primordial power
spectrum, but this is controversial (Seljak, McDonald \& Makarov 2003).

\subsection{Metals in the Diffuse IGM}

High resolution spectroscopy on 8-10m class telescopes has opened up a
new window for studying metal absorption in low-density regions of the
Universe.  Using Keck/HIRES, Songaila \& Cowie (1996) found that most
Ly$\alpha$ systems at $z\sim 3$ with $N_{\rm HI}>10^{15}{\rm cm}^{-2}$,
and half the systems with $N_{\rm HI}>10^{14.5}{\rm cm}^{-2}$, showed
associated \ion{C}{IV} absorption.  This is fairly surprising, since
the latter column density limit corresponds to matter overdensities of
a few at those redshifts, which presumably lie well outside of galaxies.
Thus some mechanism must be invoked to transport metals across 
cosmological distances.  Constraints on such mechanisms can be provided
by examining the density dependence of the metallicity, $Z(\rho)$:
A strong gradient would imply late, local enrichment, while a more 
uniform distribution would suggest an early epoch of enrichment.

There are two approaches to studying $Z(\rho)$: One is to trace a single
ion over a range of IGM densities, and the other is to examine different
ions that trace different densities.  The idea behind the latter is shown
in the left panel of Figure~3:  The curves show the Line Observability
Index (LOX; Hellsten et al. 1998) for \ion{C}{IV} and \ion{O}{VI}, as
a function of $N_{\rm HI}$ (which, via the FGPA, yields the ionization
conditions).  The curves shown use a metallicity [C/H]$=-2.5$ and a halo
star abundance pattern with [O/C]$=0.5$.  The LOX translates to an 
equivalent width, in m\AA, of a ``typical" associated metal absorber.
The solid lines show the LOX for a Haardt \& Madau (1996) background,
while the dashed lines show the effects of attenuating this background
by $\times 10$ above 4~Ryd in order to mimic the ionization conditions
prior to \ion{He}{II} reionization (e.g. Heap et al. 2000).  Notice that
a comparison of these two ions can provide not only an estimate of $Z(\rho)$,
but also constrain the shape of the ionizing flux.  The horizontal lines
denote approximate detection limits for Keck/HIRES, for within the
Ly$\alpha$ forest (dashed; appropriate for \ion{O}{VI}) and redwards
of it (solid; for \ion{C}{IV}).  From this, it is predicted that \ion{C}{IV}
would routinely be seen for $N_{\rm HI}>10^{14.5}{\rm cm}^{-2}$,
as observed, but at lower densities it is increasingly ionized to \ion{C}{V}.
Meanwhile, the lowest density gas is probed nicely with \ion{O}{VI}
absorption.  Systems with $N_{\rm HI}\approx 10^{14}{\rm cm}^{-2}$
correspond to gas at the mean density of the Universe at $z=3$, so
in principle \ion{O}{VI} absorption can even probe into voids.  Note
that \ion{O}{VI} here is photoionized, not collisionally ionized,
consistent with observations of \ion{O}{VI} absorbers at $z\ga 2$
(Carswell et al. 2002).

\begin{figure}
\plotfiddle{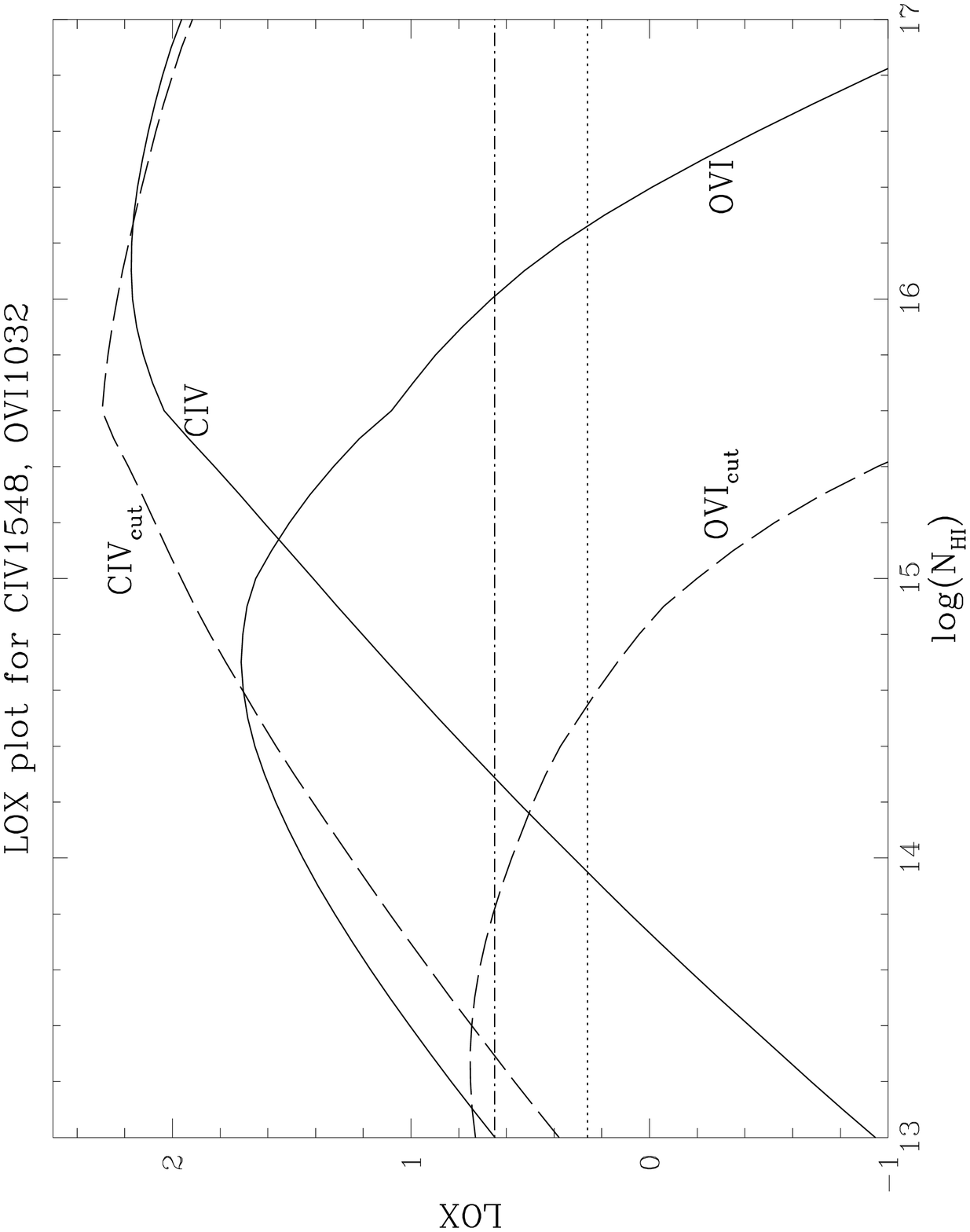}{1in}{270}{28}{28}{-220}{85}
\plotfiddle{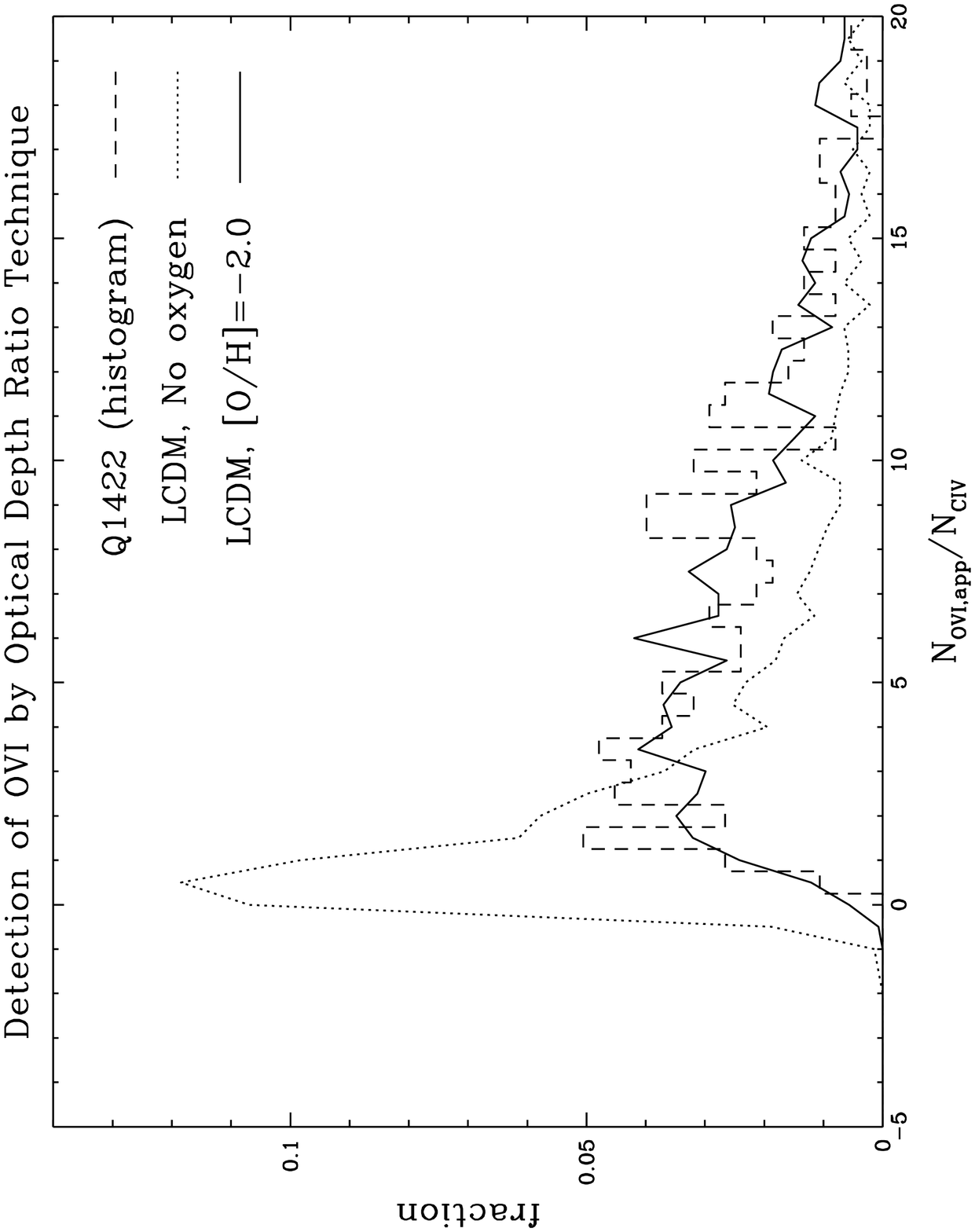}{1in}{270}{28}{28}{-10}{170}
\caption{{\it Left panel:} Line Observability Index (LOX) for \ion{C}{IV}
and \ion{O}{VI} at $z=3$, for a quasar background (solid) and a pre-\ion{He}{II}
reionization (``cut") background (dashed).
{\it Right panel:} Apparent optical depth ratios for pixels in Q1422 spectrum
with significant \ion{C}{IV} absorption.  Comparison with simulated 
spectra clearly indicates the presence of associated \ion{O}{VI} absorption.
Figures from Dav\'e et al. (1998).}
\end{figure}

The practical difficulty with using \ion{O}{VI} is that its doublet
lies at rest wavelengths 1032\&1037\AA, firmly within the dense Ly$\alpha$ forest, so it
is quite difficult to identify and confirm the doublet lines.  Instead,
Dav\'e et al. (1998) followed Songaila (1998) in taking a pixel-based
approach to identify an excess of absorption over the expected
contribution from \ion{H}{I} at the wavelength of associated \ion{O}{VI}.
The so-called optical depth ratio technique yields the plot shown in 
the right panel of Figure~3:  The histogram of apparent optical depth
ratios $N_{\rm OVI}/N_{\rm CIV}$ from a Keck/HIRES spectrum of Q1422+2309 
is compared to predictions from a hydro simulation with and without oxygen.
This shows that \ion{O}{VI} is clearly present in \ion{C}{IV} systems,
at a level consistent with a quasar-dominated, post-\ion{He}{II} recombination
ionizing background and an alpha-enhanced abundance pattern.  Dav\'e et al.
also found {\it no} \ion{O}{VI} in weaker \ion{H}{I} systems that showed
no \ion{C}{IV}.    From these results, Dav\'e et al. reached two main
conclusions:
\begin{enumerate}
\item A comparison of \ion{O}{VI} in systems with and without \ion{C}{IV}
indicates a metallicity gradient of $dZ/d\log\rho \ga 0.5$.
\item The presence of \ion{O}{VI} means that \ion{He}{II} was reionized
over at least half the redshift path length covered by Q1422's Ly$\alpha$
forest from $z=3.5\rightarrow 3.1$.
\end{enumerate}
Though much larger data sets now exist, this work has not been followed 
up due to the lack of availability of Keck/HIRES quasar spectra.

Meanwhile, VLT's UVES spectrograph has now obtained large samples of 
high-redshift quasar spectra.  Aguirre et al. (2002) developed a somewhat
different pixel-based analysis technique called the Pixel Optical Depth
(POD) method to optimally extract metallicity information in low-density 
regions.  Using 9 UVES spectra and hydro simulations for calibration, 
Schaye et al. (2003) quantified the
variation in IGM metallicity with both density and redshift, employing the
simpler, more robust idea of measuring \ion{C}{IV} absorption over
a sizeable range of \ion{H}{I} optical depths.  They obtained
[C/H]$=-3.47^{+0.06}_{-0.07} + 0.08^{+0.09}_{-0.10} (z-3) +
0.65^{+0.10}_{-0.14} (\log\delta-0.5)$, over $-0.5\la \log\delta\la 1$
and $2\la z\la 4$.  Thus they find a modest gradient 
with density (consistent with Dav\'e et al. 1998), but little evolution of 
the IGM metallicity.  A similar lack of evolution in \ion{C}{IV} was also found by 
Songaila (2001) for $z=5.5\rightarrow 2$.  Aguirre et al. (2003) extended
this method to \ion{Si}{IV}, which generally traces higher densities
than \ion{C}{IV}, and confirmed the lack of evolution in a new ion,
in addition to finding intriguing evidence for large alpha enhancements
([Si/C]$\approx 0.77$) in the IGM.  These results generally point towards
an early epoch of enrichment at $z>>3$.

Conversely, recent observations of superwinds from Lyman break ($z\sim 3$) 
galaxies (Pettini et al. 2002) argues in favor of a local enrichment 
mechanism.  Adelberger et al. (2003) find a curious {\it lack} of \ion{H}{I}
absorption in the immediate vicinity of galaxies
at impact parameters $\la 0.5$ comoving Mpc/h
(see the solid points in the left panel of Figure~4), 
which they interpret as further evidence for
strong winds.  Adelberger (these proceedings) also found a strong enrichment
of \ion{C}{IV} along lines of sight passing near galaxies, providing
direct evidence for {\it in situ} enrichment.  Thus the exact manner and
epoch at which the IGM was enriched remains uncertain.

\subsection{Absorption Near Galaxies at $z\sim 3$}

\begin{figure}
\plotfiddle{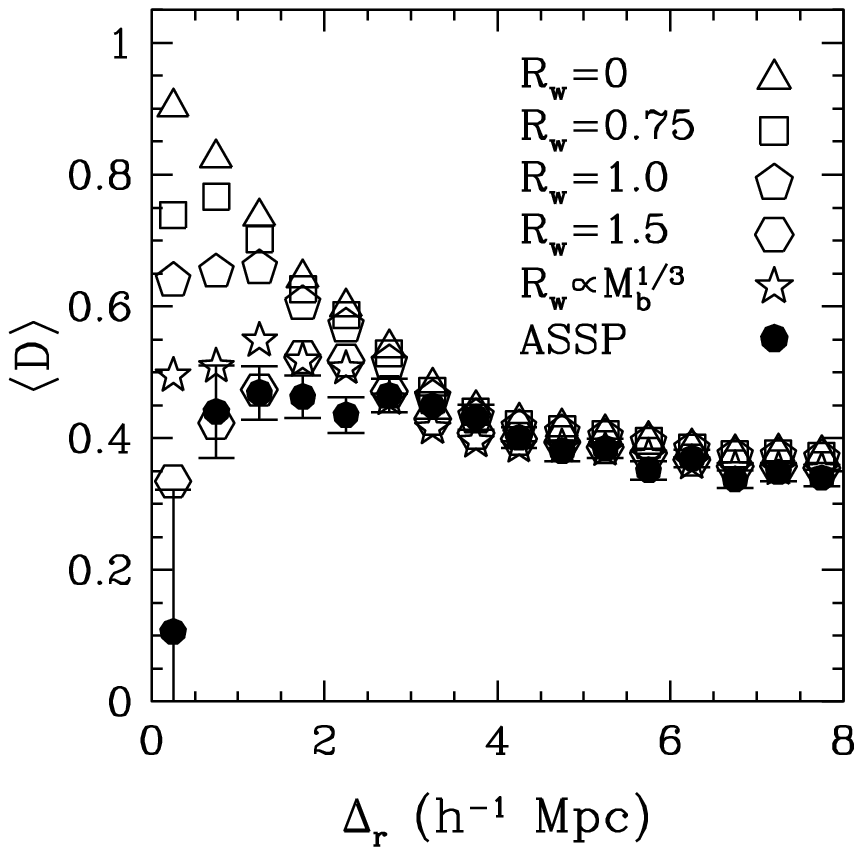}{1in}{0}{75}{75}{-240}{-450}
\plotfiddle{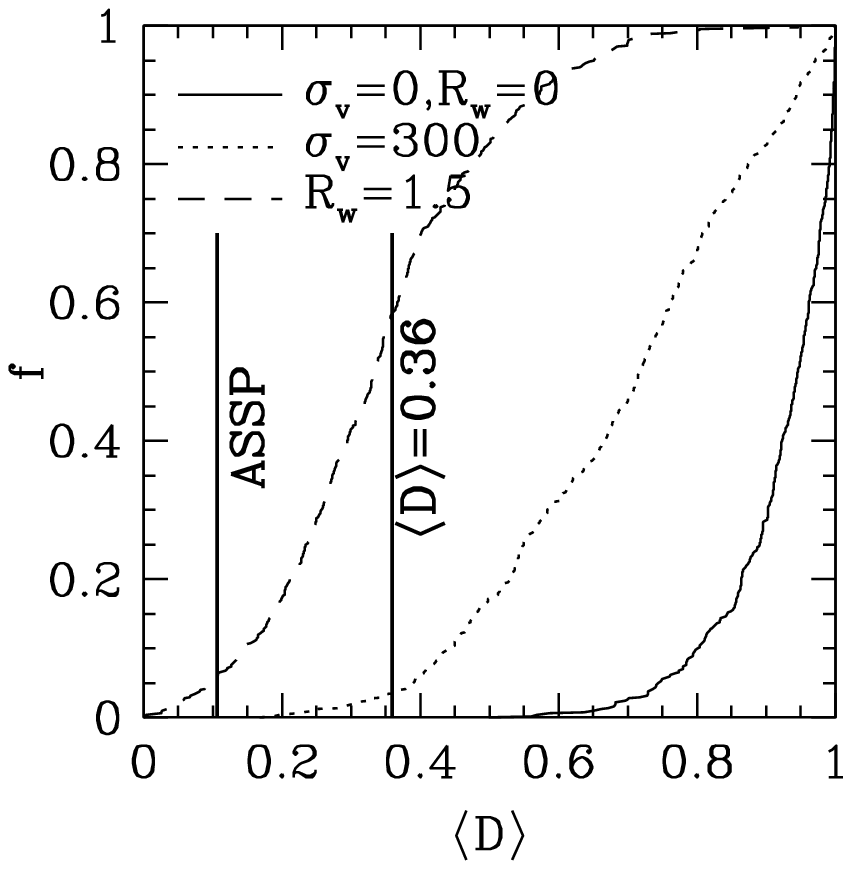}{1in}{0}{75}{75}{-50}{-365}
\caption{{\it Left panel:} Ly$\alpha$ flux decrement as a function of radius
from galaxies at $z\sim 3$ for various evacuation models described in text, 
compared with Adelberger et al. data.
{\it Right panel:} Statistical probability of reproducing the innermost Adelberger et al.
point, in various model scenarios.  Figures from Kollmeier et al. (2003).}
\end{figure}

The galaxy-IGM connection explored by Adelberger et al. (2003) also presents
a mystery.  The dramatic drop in absorption near galaxies is opposite to
what is expected from hierarchical simulations, since matter clustering tends
to increase absorption close to galaxies (Kollmeier et al. 2003).  The
left panel of Figure~4 shows the Adelberger et al. data (labelled ``ASSP")
compared to a hydro simulation having no winds/photoionization (triangles),
and various {\it ad hoc} model with complete evacuation of \ion{H}{I}
out to $R_w=0.75$ (squares), 1 (pentagons), and 1.5 (hexagons) comoving
Mpc/h, and another model where the evacuation volume is proportional to
the galaxies' baryonic mass (stars).  Only in the extreme case of complete
evacuation out to 1.5 Mpc/h does the model come within the error bars of
the data.  The radius would have to be even larger if the evacuation was
(more realistically) non-spherical or incomplete.  Large evacuation radii are
required because peculiar velocities can shift absorption from large distances
to the galaxies' redshifts (Desjacques et al. 2003).  Models 
based on local photoionization by 
star formation fail by a considerable margin to produce the required evacuations,
so winds are the logical explanation, but the required energies are
quite large.  Similar conclusions were found by Croft et al. (2002).

The fact that only three quasar-galaxy pairs contribute to Adelberger et al.'s
innermost point raises the possibility that the lack of absorption is
a statistical fluke.  To examine this, Kollmeier et al. sampled trios of
galaxy-absorber pairs in various models to determine the frequency 
of obtaining a flux decrement as low as observed.  Figure~4, right panel,
shows that it is exceptionally rare for our fiducial simulation, even
when a random redshift error of $\sigma_v=300$~km/s is included.  Even
for our most extreme evacuation model there is barely a 5\% chance of
reproducing the data.

It remains unclear what physics is required to reconcile the models with
this data.  Kollmeier et al. considered photoionization from
AGNs, ``filling in " of Ly$\alpha$ absorption from cooling radiation
(Fardal et al. 2001), non-equilibrium effects, and other ideas, but 
none appeared promising.  Obviously a larger sample of these systems
would be helpful, and has already been obtained at $z\sim 2$ (Steidel, priv.
comm.).  If these observational results are
borne out, it may require a radical change in our understanding of
feedback processes.

\section{The Low-Redshift Intergalactic Medium}

\subsection{The Ly$\alpha$ Forest at $z\la 1.5$}

The launch of {\it HST} with its ultraviolet capabilities heralded 
a new age in IGM studies, as the Ly$\alpha$
absorption line (and various UV metal features) became routinely observable for systems with $z\la 1.5$.
The Quasar Absorption Line Key Project compiled over 80 spectra (Jannuzi 
et al. 1998), and quantified the statistics of low-$z$ absorbers with
great precision (Weymann et al. 1998), even though the Faint Object 
Spectrograph (FOS) data could not resolve individual Ly$\alpha$ absorbers.
A major surprise was that any absorbers were found at all, since an
extrapolation from the rapidly-declining population at $z\sim 4\rightarrow 2$
would predict virtually none (Morris et al. 1991).

Dav\'e et al. (1999) extended hydrodynamic simulations to $z=0$ using
the first distributed-memory parallel TreeSPH code (Dav\'e, Dubinksi,
\& Hernquist 1997), and found that the statistical properties
of simulated absorbers were in good agreement with Key Project results.
Furthermore, they argued against a second population of slowly-evolving
absorbers dominating at low-$z$ (Bahcall et al. 1996), and instead claimed
that the change in evolution of the absorber population was caused by
a drop in the amplitude of the quasar-dominated metagalactic flux, which
increased the neutral fraction and balanced the drop in gas column density 
due to cosmic expansion.  In other words, low-$z$ Ly$\alpha$ absorbers,
at least the weaker ones, are physically analogous to their high-$z$
counterparts, albeit shifted to somewhat lower column densities.

\begin{figure}
\plotfiddle{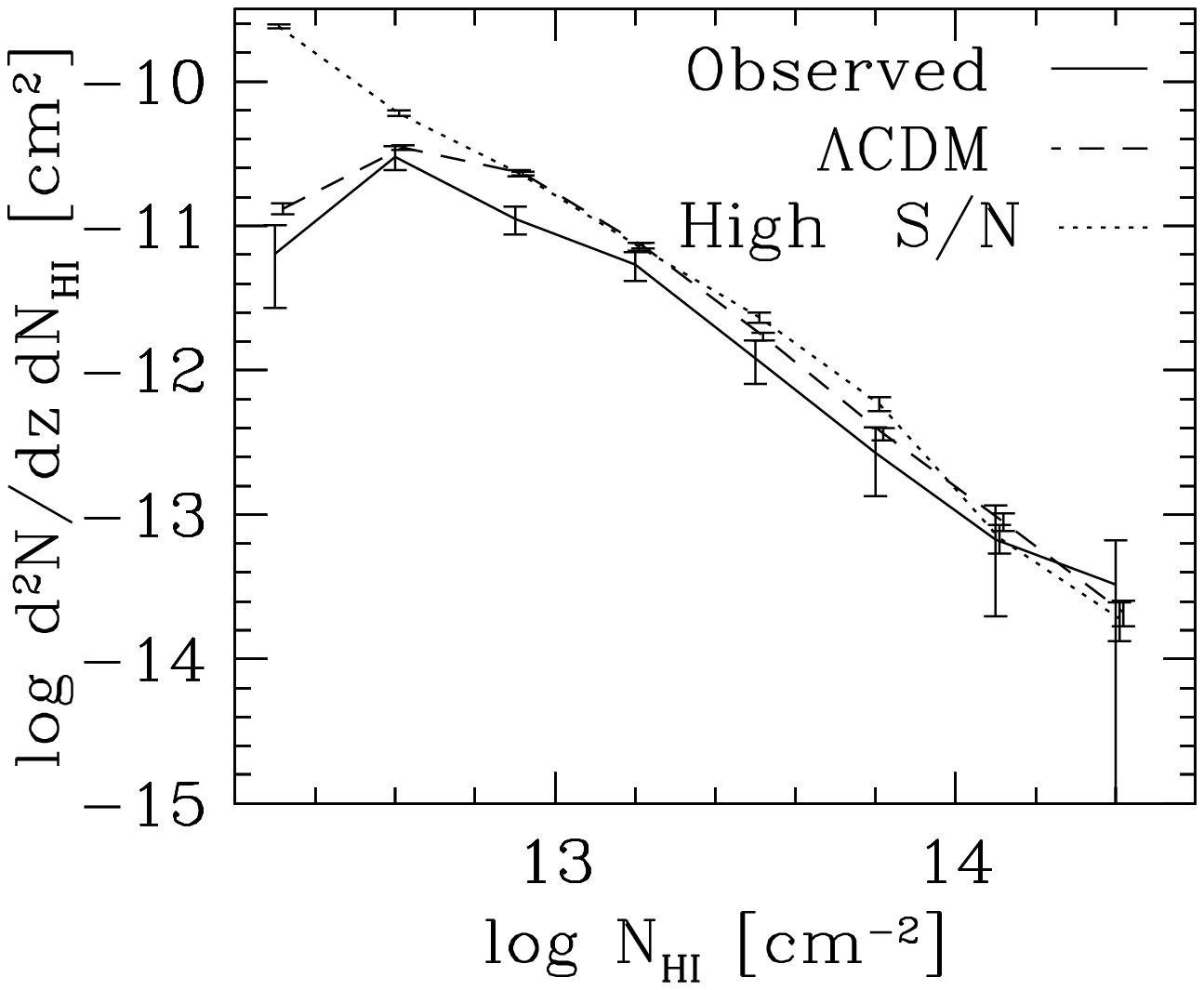}{1in}{0}{50}{50}{-240}{-280}
\plotfiddle{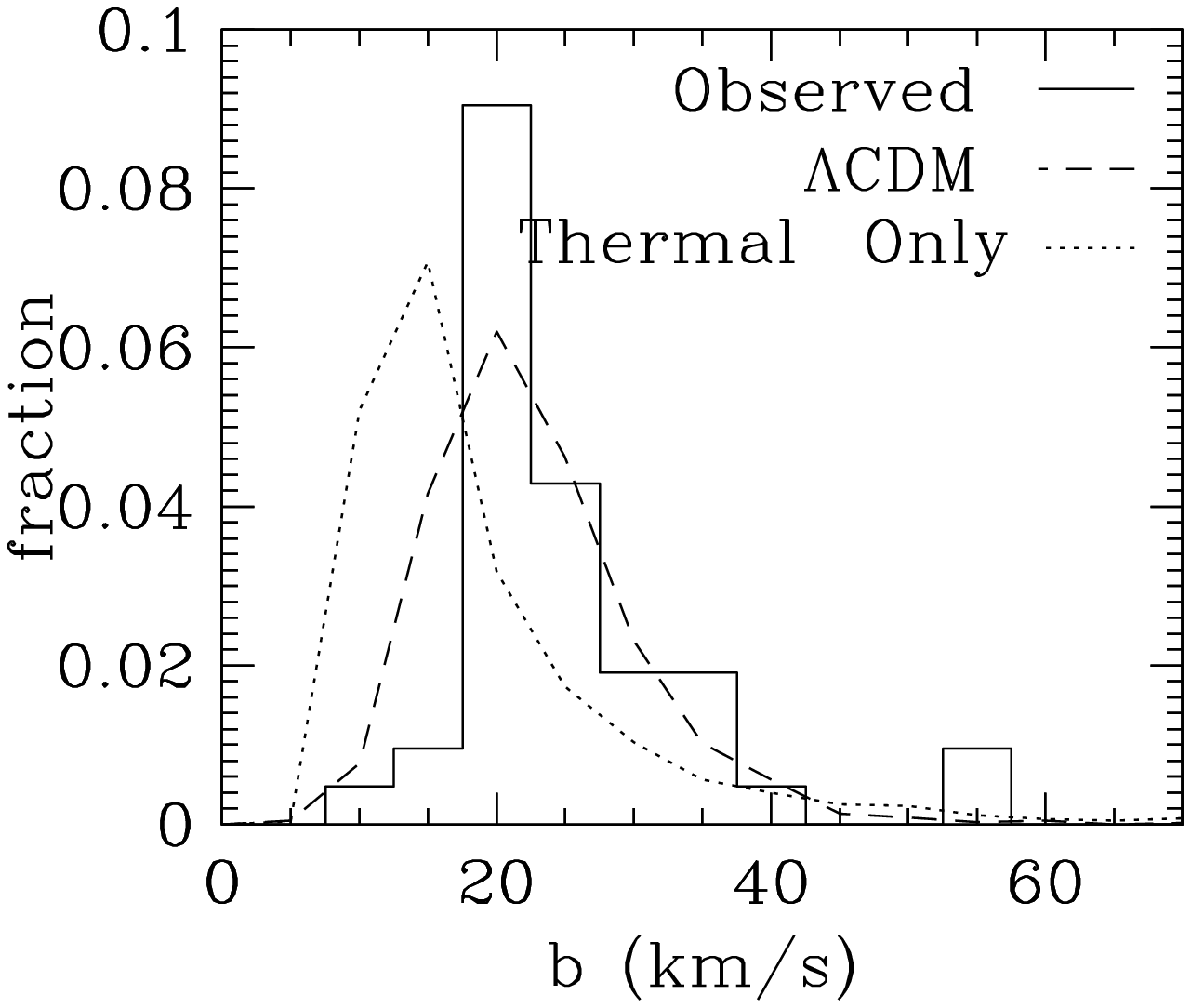}{1in}{0}{50}{50}{-50}{-195}
\caption{{\it Left panel:} Column density distribution of Ly$\alpha$ 
absorbers at $\bar z=0.17$ from PG0953+415 and H1821+643 (solid), vs. 
$\Lambda$CDM simulation (dashed).  Turnover is due to noise; higher-S/N
simulated spectra (dotted) show no break in power law.
{\it Right panel:} Linewidth ($b$-parameter) distributions.  Thermal
contribution to linewidths (dotted) is significant at low-$z$.
Figures from Dav\'e \& Tripp (2001).}
\end{figure}

The Space Telescope Imaging Spectrograph (STIS), installed in 1997, provided
the first fully resolved glimpse at the local Ly$\alpha$ forest.  Careful
comparisons versus hydro simulations continued to show excellent agreement
with data (Dav\'e \& Tripp 2001; see Figure~5).  This suggests that with
a large sample of weak absorbers, as should be provided by the Cosmic
Origins Spectrograph beginning in 2005, many of the analyses done
at high-$z$ could be performed at $z\la 1.5$.  Hence COS offers great promise
for constraining the metagalactic ionizing flux (Dav\'e \& Tripp 2001),
the IGM metallicity, and possibly even the matter power spectrum at $z\la 1$
on mildly nonlinear scales.

\subsection{The Warm-Hot Intergalactic Medium}

Despite the physical similarities in Ly$\alpha$ absorbers, the low-$z$
IGM is considerably more complex than the high-$z$ one.  As shown in 
Figure~1, at $z>3$ the diffuse IGM contains most of the baryons
in the Unvierse, but this is not true at low redshifts.  In particular,
the elusive Warm-Hot Intergalactic Medium (WHIM) is predicted to contain
a third to half of all baryons today (Dav\'e et al. 2001).

The WHIM has drawn widespread interest recently because it may be the
repository of the so-called ``missing baryons".  Fukugita, Hogan \& Peebles (1998)
performed an inventory of baryons observed today and found roughly a half
to be missing, compared with high-$z$ Ly$\alpha$ forest measurements
or the now well-determined $\Omega_b$.  Cen \& Ostriker (1999) suggested
that they reside in a warm-hot state, with $10^5<T<10^7$K, making them
difficult to detect.  Dav\'e et al. (2001) examined the physical properties
of WHIM gas in more detail, finding that it generally resides in filamentary
large-scale structures, well outside of virialized objects (Figure~6,
left panel).  The existence of such gas in large quantities was at first
surprising, since if it is all placed in virialized halos then the soft
X-ray background would be overproduced by orders of magnitude (Pen 1999;
Wu, Fabian \& Nulsen 1999).  Instead, the low densities predicted by
simulations result in a much weaker background, broadly consistent
with observational limits (Croft et al. 2001; Phillips, Ostriker \& Cen 2001).

\begin{figure}
\plottwo{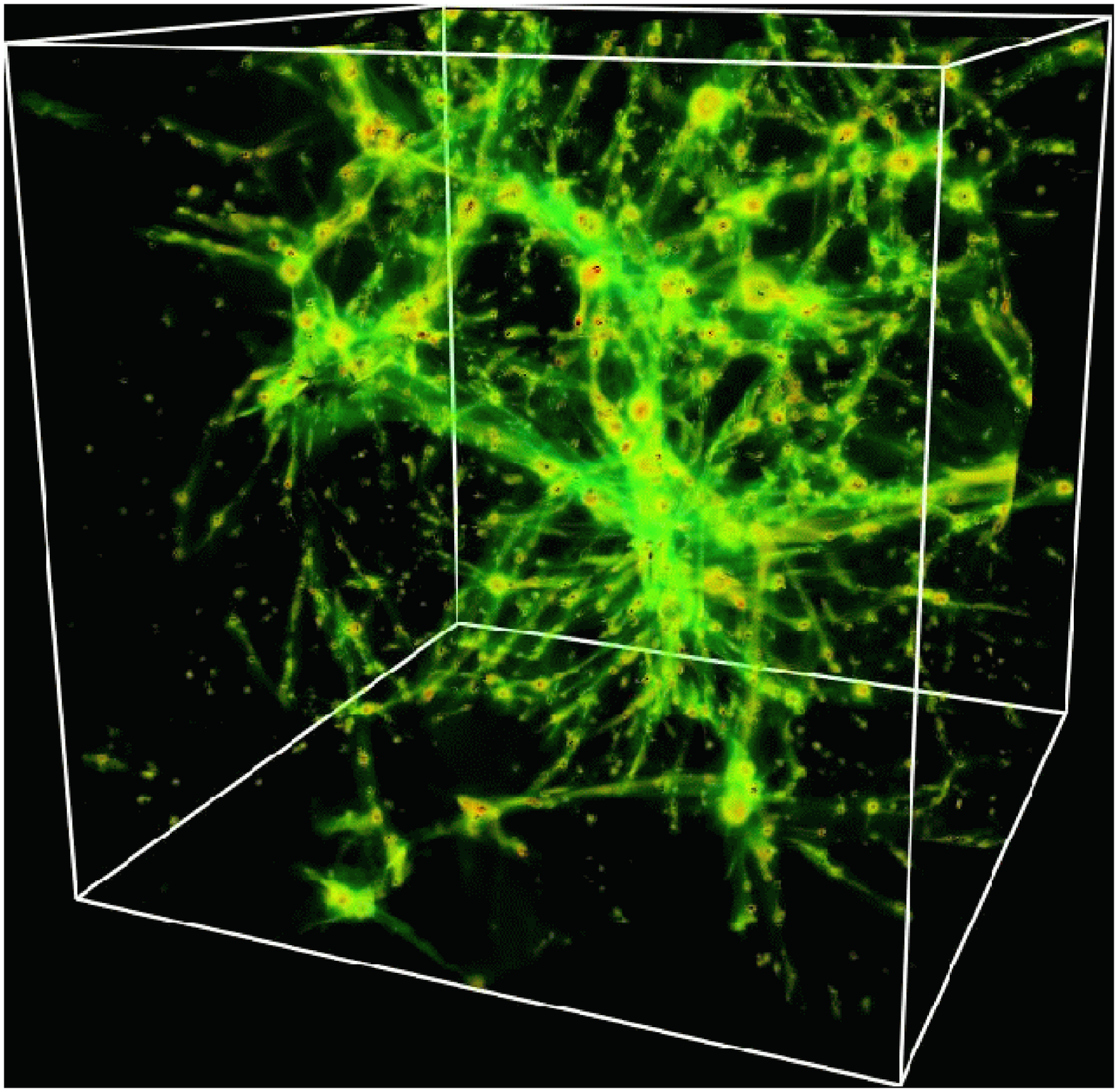}{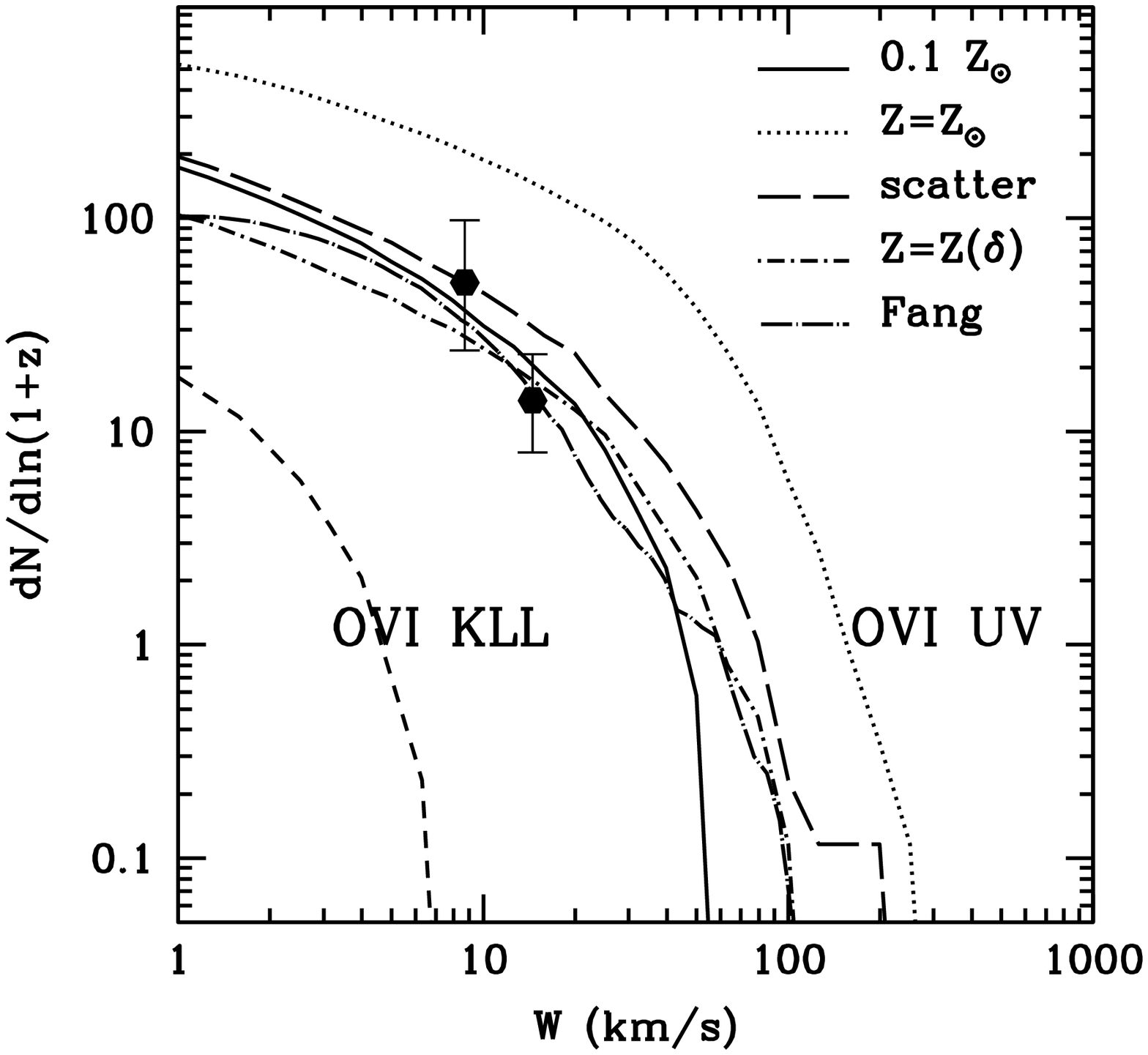}
\caption{{\it Left panel:} The Warm-Hot Intergalactic Medium in a simulation
by Cen \& Ostriker.  Green through red shows $T_{\rm gas}=10^5\rightarrow 10^7$K.
Figure from Dav\'e et al. (2001).
{\it Right panel:} Line density of \ion{O}{VI} absorbers seen in a
cosmological hydro simulation (lines) with various assumption regarding
how metals trace density.  Data points show observations of Tripp \& collaborators.
Figure from Chen et al. (2003).}
\end{figure}

Detecting the WHIM in absorption is challenging, since \ion{H}{I} is a poor
tracer of gas at these temperatures.  However, \ion{O}{VI} is promising
because it has a collisional ionization maximum at $\approx 3\times 10^5$K.
Tripp, Savage \& Jenkins (2000) detected four intervening \ion{O}{VI}
systems at $z\la 0.3$, and obtained a lower limit on the baryon density
contained in such systems to be $\ga 0.1\Omega_b$.  Because of the 
uncertainties in ionization corrections, metallicities, and gas
densities, it is perhaps more straightforward to make predictions
from simulations regarding the expected frequency of \ion{O}{VI}
absorbers.  The right panel of Figure~6 shows such a comparison from
Chen et al. (2003).  Observations (solid points) are reproduced nicely
with a metallicity of roughly one-tenth solar for WHIM gas, a metallicity
that is somewhat higher than Schaye et al. (see \S4.3) at WHIM densities
($\delta\sim 30$), extrapolated to $z\sim 0$.
Cen et al. (2001) and Fang \& Bryan (2002) obtain similar results.
\ion{O}{VI} absorption has also been detected in association with the
Local Group (Nicastro et al. 2003).

It is also possible to detect WHIM absorption in the X-ray band, particularly
using the \ion{O}{VII} and \ion{OVIII} lines (Chen et al. 2003; Fang, Bryan
\& Canizares 2003).  Mathur, Weinberg \& Chen (2003) claim to have
detected WHIM \ion{O}{VII} absorption in a {\it Chandra} spectrum of
H1821, though {\it XMM} follow-up did not unambiguously confirm it.

Detecting WHIM emission, while even more difficult, may ultimately prove
the best approach for obtaining a complete map of missing baryons in
the local Universe.  Zappacosta et al. (2002) found a diffuse soft X-ray
structure associated with a galaxy overdensity at $z=0.45$, but a more
systematic search will have to await new satellites.  The proposed
Missing Baryon Explorer (MBE) would detect a significant number of WHIM
sources in six X-ray emission lines (Fang et al. 2003).  The Japanese
have a similar proposal called Diffuse Intergalactic Oxygen Surveyor (DIOS).
Emission in lower ionic species such as \ion{O}{VI} and \ion{C}{V} would
detect the bulk of WHIM gas at lower temperatures, and the proposed Spectroscopy and Photometry
of the IGM's Diffuse Radiation (SPIDR) satellite was aiming to do
so until it was cancelled in the design phase.

\section{Summary}

Observations of the IGM have progressed rapidly in recent years, which in
turn has spurred dramatic advances in our theoretical understanding of the
various baryonic components in the Universe.  Still, many questions
crucial to our understanding of galaxy and structure formation
remain unanswered, including:
\begin{itemize}
\item What are the sources that reionize the IGM, and how can we detect them?
\item How and when did metals get into the IGM?
\item What effects do galactic feedback have on surrounding IGM gas, and how
does it change with redshift?
\item How do we obtain a complete observational census of baryons today?
\end{itemize}
Here I have given a status report on these
topics, which currently present more questions than answers.  Major advances
will only come through joint progress in observational
capabilities, particularly in the ultraviolet, along with improving
simulations for the evolution of baryons in the Universe.

\acknowledgements

I wish to thank the various collaborators on projects mentioned in these
proceedings, including Betsy Barton, Greg Bryan, Renyue Cen, Xuelei Chen,
Rupert Croft, John Dubinski, Taotao Fang, Mark Fardal, Uffe Hellsten, Lars Hernquist, Neal Katz, Juna
Kollmeier, Jerry Ostriker, Casey Papovich, J.-D. Smith, Volker Springel,
Todd Tripp, and David Weinberg.


\begin{references}

\reference Abel, T., Bryan, G. L., \& Norman, M. L. 2002, Science, 295, 93
\reference Adelberger, K. L., Steidel, C., Shapley, A., \& Pettini, M. 2003, \apj, 584, 45
\reference Aguirre, A., Schaye, J., Kim, T.-S., Theuns, T., Rauch, M., \& Sargent, W. L. W. 2003, \apj, in press, astro-ph/0310664
\reference Aguirre, A., Schaye, J., Theuns, T. 2002, \apj, 576, 1
\reference Bahcall, J. N. \& Salpeter, E. E. 1965, \apj, 142, 1677
\reference Barton, E. J., Dav\'e, R., Smith, J.-D. T., Papovich, C., Hernquist, L., \& Springel, V. 2003, \apjl, submitted, astro-ph/0310514
\reference Black, J. 1981, \mnras, 197, 533
\reference Bromm, V., Coppi, P. S., \& Larson, R. B. 2002, \apj, 564, 23
\reference Carswell, R., Schaye, J., \& Kim, T.-S. 2002, \apj, 578, 43
\reference Cen, R. 2003, \apj, 591, 12
\reference Cen, R. 2003b, \apjl, 591, L5
\reference Cen, R., Miralda-Escud\'e, J., Ostriker, J.P., \& Rauch M. 1994, \apj, 427, L9
\reference Cen, R. \& Ostriker, J. P. 1999, \apj, 514, 1
\reference Cen, R., Tripp, T. M., Ostriker, J. P., \& Jenkins, E. B. 2001, \apj, 559, L5
\reference Chen, X., Weinberg, D. H., Katz, N., \& Dav\'e, R. 2003, ApJ, 594, 42
\reference Choudhury, T. R., Padmanabhan, T., Srianand, R. 2001, \mnras, 322, 561
\reference Ciardi, B. \& Madau, P. 2003, \apj, 596, 1
\reference Croft, R. A. C., Hernquist, L., Springel, V., Westover, M., \& White, M. 2002, \apj, 580, 634
\reference Croft, R. A. C., Weinberg, D. H., Bolte, M., Burles, S., Hernquist, L., Katz, N., Kirkman, D., \& Tytler, D. 2002, \apj, 581, 20
\reference Croft, R. A. C., Weinberg, D. H., Katz, N., \& Hernquist, L. 1998, \apj, 495, 44
\reference Dav\'e, R., Dubinski, J., \& Hernquist, L. 1997, NewAst, 2, 277
\reference Dav\'e, R., Hernquist, L., Weinberg, D. H., \& Katz, N. 1997, \apj, 477, 21
\reference Dav\'e, R., Hellsten, U., Hernquist, L., Katz, N., \& Weinberg, D. H. 1998, ApJ, 509, 661.
\reference Dav\'e, R., Hernquist, L., Katz, N., \& Weinberg, D. H. 1999, ApJ, 511, 521
\reference Dav\'e, R., Cen, R., Ostriker, J. P., Bryan, G. L., Hernquist, L., Katz, N., Weinberg, D. H., Norman, M. L., \& O'Shea, B. 2001, ApJ, 552, 473
\reference Dav\'e, R. \& Tripp, T. M. 2001, \apj, 553, 528
\reference Desjacques, V., Nusser, A., Haehnelt, M. G., \& Stoehr, F. 2003, \mnras, submitted, astro-ph/0311209
\reference Fang, T. \& Bryan, G. L. 2001, \apj, 561, L31
\reference Fang, T., Bryan, G. L., Canizares, C. R. 2002, \apj, 564, 604
\reference Fang, T., Croft, R. A. C., Sanders, W. T., Houck, J., Dav\'e, R., Katz, N., Weinberg, D. H., Hernquist, L. 2003, \apj, submitted, astro-ph/0311141
\reference Fan, X. et al. 2003, \aj, 125, 1649
\reference Fardal, M. A., Katz, N., Gardner, J. P., Hernquist, L., Weinberg, D. H., \& Dav\'e, R. 2001, \apj, 562, 605
\reference Fukugita, M., Hogan, C. J., \& Peebles, P. J. E. 1998, \apj, 503, 518
\reference Furlanetto, S. R. \& Loeb, A. 2002, \apj, 579, 1
\reference Furlanetto, S. R., Sokasian, A., \& Hernquist, L. 2003, \mnras, submitted, astro-ph/0305065
\reference Gunn, J. E. \& Peterson, B. A. 1965, \apj, 142, 1633
\reference Haardt, F. \& Madau, P. 1996, \apj, 461, 20
\reference Haiman, Z. 2002, \apj, 576, L1
\reference Haiman, Z., Abel, T. \& Rees, M. J. 2000, \apj, 534, 11
\reference Heap, S. R., Williger, G. M., Smette, A., Hubeny, I., Sahu, M. S., Jenkins, E. B., Tripp, T. M., \& Winkler, J. N. 2000, \apj, 534
\reference Heger, A. \& Woosley, S. E. 2002, \apj, 567, 532
\reference Hellsten, U., Hernquist, L., Katz, N., \& Weinberg, D. H. 1998, \apj, 499, 172
\reference Hernquist, L.H., Katz, N., Weinberg, D.H., \& Miralda-Escud\'e, J. 1996, \apjl, 457, L51
\reference Hui, L. \& Gnedin, N. Y. 1997, \mnras, 292, 27
\reference Ikeuchi, S. 1986, Astrophys. Space Sci. 118, 509
\reference Ikeuchi, S. \& Ostriker, J. P. 1986, \apj, 301, 522
\reference Iliev, I. T., Shapiro, P. R., Ferrara, A., \& Martel, H. 2002, \apjl, 572, L123
\reference Iliev, I. T., Scannapieco, E., Martel, H., \& Shapiro, P. R. 2003, \mnras, 341, 81
\reference Jannuzi, B. T. et al. 1998, \apjs, 118, 1
\reference Kogut, A. et al. 2003, \apjs, 148, 161
\reference Kollmeier, J. A., Weinberg, D. H., Dav\'e, R., \& Katz, N. 2003, ApJ, 594, 75
\reference Lynds C. R. \& Stockton A. N. 1966, \apj, 144, 446
\reference Madau, P., Rees, M. J., Volonteri, M., Haardt, F., \& Oh, S. P. 2003, \apj, submitted, astro-ph/0310223
\reference Melott, A. 1980, \aj, 268, 630
\reference Morris, S. L., Weymann, R. J., Savage, B., Gilliland, R. 1991, \apj, 377, L21
\reference Nicastro, F., Zezas, A., Elvis, M., Mathur, S., Fiore, F., Cecchi-Pestellini, C., Burke, D., Drake, J., \& Casella, P. 2003, Nature, 421, 719
\reference Oh, S. P. \& Haiman, Z. 2003, \mnras, submitted, astro-ph/0307135
\reference Oh, S. P., Nollett, K. M., Madau, P., \& Wasserburg, G. J. 2001, \apjl, 562, L1
\reference Pen, U.-L. 1999, \apjl, 510, L1
\reference Petitjean, P., Bergeron, J., Carswell, R. F., \& Puget, J. L. 1993, 260, 67
\reference Pettini, M., Rix, S. A., Steidel, C. C., Adelberger, K. L., Hunt, M. P., \& Shapley, A. E. 2002, \apj, 569, 742
\reference Phillips, L. A., Ostriker, J. P., Cen, R. 2001, \apj, 554, L9
\reference Rauch, M. et al. 1997, \apj, 489, 7
\reference Rees, M. J. 1986, \mnras, 218, 25
\reference Santos, M. R. 2003, \mnras, submitted, astro-ph/0308196
\reference Santos, M. R., Bromm, V., \& Kamionkowski, M. 2002, \mnras, 336, 1082
\reference Sargent, W. L. W., Young, P. J., Boksenberg, A. \& Tytler, D. A. 1980, \apjs, 42, 41
\reference Schaye, J., Theuns, T., Rauch, M., Efstathiou, G., Sargent, W. L. W. 2000, \mnras, 318, 817
\reference Schaye, J., Aguirre, A., Kim, T.-S., Theuns, T., Rauch, M., \& Sargent, W. L. W. 2003, \apj, 596, 768
\reference Schmidt, M. 1965, \apj, 141, 1295
\reference Schneider, R., Ferrara, A., Salvaterra, R., Omukai, K., \& Bromm, V. 2003, Nature, 422, 869
\reference Seljak, U., McDonald, P., \& Makarov, A. 2003, \mnras, 342, L79
\reference Shapiro, P. R., Iliev, I. T., \& Raga, A. C. 2003, \mnras, submitted
\reference Songaila, A. 1998, \aj, 115, 2184
\reference Songaila, A. 2001, \apj, 561, L153
\reference Songaila, A. \& Cowie, L. L. 1996, \aj, 112, 335
\reference Spergel, D. N. et al. 2003, \apjs, 148, 175
\reference Springel, V. \& Hernquist, L. 2003, \mnras, 339, 312
\reference Tegmark, M., Silk, J., Rees, M. J., Blanchard, A., Abel, T., \& Palla, F. 1997, \apj, 474, 1
\reference Tytler, D., Fan, X.-M., \& Burles, S. 1996, Nature, 381, 207
\reference Valageas, P., Schaeffer, R., \& Silk, J. 2002, A\&A, 388, 741
\reference Venkatesan, A., Giroux, M. L., \& Shull J. M. 2001, \apj, 563, 1
\reference Weymann, R. J. et al. 1998, \apj, 506, 1
\reference White, R. L., Becker, R. H., Fan, X., \& Strauss, M. A. 2003, \aj, 126, 1
\reference Wu, K. K. S., Fabian, A. C., \& Nulsen, P. E. J. 2001, \mnras, 95
\reference Wyithe, J. S. B. \& Loeb, A. 2003a, \apj, 586, 693
\reference Wyithe, J. S. B. \& Loeb, A. 2003b, \apjl, 588, L69
\reference Zappacosta, L., Mannucci, F., Maiolino, R., Gilli, R., Ferrara, A., Finoguenov, A., Nagar, N. M., Axon, D. J. 2002, A\&A, 394, 7
\reference Zhang, Y., Anninos, P., \& Norman, M.L. 1995, \apjl, 453, L57
\end{references}
\end{document}